\documentclass[journal]{IEEEtran}
\usepackage{amsfonts,amssymb}
\usepackage[cmex10]{amsmath}
\usepackage{epsfig}
\usepackage{amsfonts}
\usepackage{graphics}
\usepackage{graphicx}
\usepackage{subfig}
\usepackage{mathtools}
\usepackage{cite}
\usepackage{multirow}
\epsfverbosetrue

    \def\independenT#1#2{\mathrel{\setbox0\hbox{$#1#2$}%
    \copy0\kern-\wd0\mkern4mu\box0}}

\newcommand{\el}{{ ~~\it Example: \ }}

\title{On the Stability Region of Multi-Queue Multi-Server Queueing Systems with Stationary Channel Distribution}  
\author{Hassan Halabian, Ioannis Lambadaris, Chung-Horng Lung
\\
Department of Systems and Computer Engineering, Carleton University, Ottawa, ON, Canada\\
Email: \{hassanh, ioannis, chlung\}@sce.carleton.ca
\vspace{-1cm}}      
\begin{document}             
\maketitle   
\thispagestyle{empty}                
\newtheorem{theor}{Theorem}
\newtheorem{lem}{Lemma}
\newtheorem{pro}{Proposition}
\newtheorem{defin}{Definition}
\newtheorem{conj}{Conjecture}
\newtheorem{cor}{Corollary}

\IEEEpeerreviewmaketitle
\begin{abstract}
In this paper, we characterize the stability region of multi-queue multi-server (MQMS) queueing systems with stationary channel and packet arrival processes. Toward this, the necessary and sufficient conditions for the stability of the system are derived under general arrival processes with finite first and second moments. We show that when the arrival processes are stationary, the stability region form is a polytope for which we explicitly find the coefficients of the linear inequalities which characterize the stability region polytope. 

%
\end{abstract}
\vspace{-3mm}
\section{Introduction}
\textit{Stability region} or \textit{network capacity region} of a communication network is the closure of the set of all arrival rate vectors for which there exists a resource allocation policy that can stabilize the system \cite{Now, phd, Tassiulas92}. Having the stability region of a network characterized by a set of convex inequalities, we can obtain a precise solution for network fairness optimization problems with the general form of (\ref{fairness}).
\begin{eqnarray}
\label{fairness}~~ \texttt{Maximize:}~~~~~~~~~~\sum_{n=1}^N f_n(r_n)~~~~~~~~~~~~~~~~~~~~  \\ 
\texttt{Subject to:}~~~~~~~~~~r=(r_1,r_2,...,r_N) \in \Lambda ~~~~\nonumber \\
0 \leq r_n \leq \lambda_n~~\forall n=1,2,...,N \nonumber\hspace{-.5cm}
\end{eqnarray}
In (\ref{fairness}), $N$ is the number of traffic sources. By $r=(r_1,r_2,...,r_N)$, we denote the admitted rate vector, $\lambda_n$ is the input rate of each source $n$ and $\Lambda$ denotes the stability region.
Functions $f_n(r_n)$ are non-decreasing and concave utility functions. The choice of $f_n(r_n)$ determine the desired fairness properties of the network. 
The goal of this paper is to characterize the network capacity region of Multi-Queue Multi-Server (MQMS) queueing systems with stationary channel distribution (i.e., $\Lambda$ for MQMS system). Such queueing systems are used to model multiuser wireless networks with multiple orthogonal subchannels such as OFDMA access networks. 
%
The resource allocation in such networks can be modelled as the server allocation problem in multi-queue multi-server queueing systems with time varying channel conditions \cite{javidi,javidi2,javidi4, hussein, ganti,khodam}. 

The stability problem in wireless queueing networks was mainly addressed in \cite{Tassiulas92, Tassiulas93,Now, phd,khodam,javidi2,power}. In \cite{Tassiulas92}, the authors introduced the notion of capacity region of a queueing network. They considered a time slotted system and assumed that arrival processes are i.i.d. sequences and the queue length process is a Markov process. In \cite{Tassiulas93}, they characterized the network capacity region of multi-queue single-server system with time varying ON-OFF connectivities. They also proved that for a symmetric system i.e., with the same arrival and connectivity statistics for all the queues, LCQ (Longest Connected Queue) policy maximizes the capacity region and also provides the optimal performance in terms of average queue occupancy (or equivalently average queueing delay). In \cite{Now,phd} and \cite{power}, the notion of network capacity region of a wireless network was introduced for more general arrival and queue length processes. 

The problem of server allocation in multi-queue multi-server systems with time varying connectivities was mainly addressed in \cite{javidi,javidi2,javidi4, hussein, khodam}. In \cite{javidi2}, Maximum Weight (MW) policy was proposed as a throughput optimal server allocation policy for MQMS systems with stationary channel processes. However, in \cite{javidi2} the conditions on the arrival traffic to guarantee the stability of MW were not explicitly mentioned. 
In our previous work \cite{khodam}, we characterized the network capacity region of multi-queue multi-server systems with time varying ON-OFF channels. We also obtained an upper bound for the average queueing delay of AS/LCQ (Any Server/Longest Connected Queue) policy which is the throughput optimal server allocation policy for such systems.


In this paper, we characterize the stability region of multi-queue multi-server queueing system with stationary channel and arrival processes. Toward this, the necessary and sufficient conditions for the stability of the system are derived for general arrival processes with finite first and second moments. For stationary arrival processes, these conditions establish the network stability region of such systems. 
Our contribution in this work is to characterize the stability region as a \textit{convex polytope} specified by a finite set of linear inequalities that can be numerically tabulated and used to solve network optimization problem (\ref{fairness}).
%
We further introduce an upper bound for the average queueing delay of MW policy which is a throughput optimal policy for MQMS systems. 

The rest of this paper is organized as follows. Section \ref{modeldes} describes the model and notation required throughout the paper. In section \ref{back}, we discuss about some preliminary definitions (stability definition, polytope, etc.) used in the paper. In section \ref{sec-asli}, we will derive necessary and sufficient conditions for the stability of our model and also find an upper bound for the average queue occupancy (or average queueing delay). Section \ref{conc} summarizes the conclusions of the paper. 
\vspace{-2mm}
\section {Model Description} \label{modeldes}
\begin{figure}[tp]
    \centering
    \includegraphics[width=0.35\textwidth]{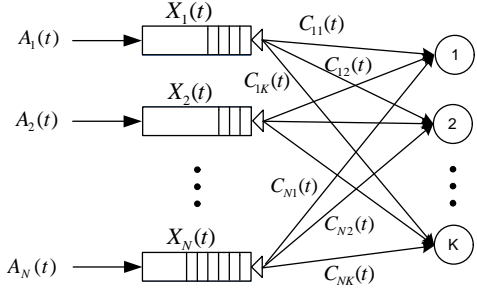}\vspace{-1mm}
	\caption{ Multi-queue multi-server queueing system with stationary channel distribution}
	\label{model}\vspace{-6mm}
\end{figure} 
Before we proceed to describe the model, we introduce basic notations often used throughout the paper. Other notations are introduced whenever it is necessary. Through this paper, all the vectors are considered to be row vectors. The time average of a function $f(t)$ is denoted by $ \overline{f(t)} $, i.e.,  $\overline{f(t)}= \lim_{t\rightarrow \infty}\frac{1}{t} \sum_{\tau=1}^{t} f(\tau)$. The operator $``\circledast"$ is used for entry-wise multiplication of two matrices. By $\underline{1}_K$ ($\underline{0}_K$), we denote a row vector of size $K$ whose elements are all identically equal to $``1"$ ($``0"$). The cardinality of a set is shown by $|\cdot|$. 
 For any vector $\alpha=(\alpha_1,\alpha_2,...,\alpha_N)$ and a non-empty index set $ \mathcal{U}=\{u_1,u_2,...,u_{|\mathcal{U}|}\} \subseteq \{1,2,...,N\} $ and $u_1<u_2<...< u_{|\mathcal{U}|}$, we define $ \alpha_{\mathcal{U}}=(\alpha_{u_1},\alpha_{u_2},...,\alpha_{u_{|\mathcal{U}|}}) $.

 We consider a time slotted queueing system with equal length time slots and equal length packets. The model consists of a set of parallel queues ${\cal N}=\{1,2,...,N\}$ and a set of identical servers ${\cal K}=\{1,2,...,K\}$ (Figure \ref{model}). Each server can serve at most one queue at each time slot, i.e., we do not have server sharing in the system. At each time slot $t$, the capacity of each link between each queue $ n\in \mathcal{N}$ and server $k\in \mathcal{K}$ is assumed to be $C_{n,k}(t)$ packets/slot. We assume that $C_{n,k}(t) \in \mathcal{M}$ where $\mathcal{M}=\{m\in \mathbb{Z}_+ \mid m \leq M\} $ for given $M$. Therefore, at each time slot $t$, the channel state may be expressed by an $N\times K$ matrix of $C(t)=\left( C_{n,k}(t)\right),n\in \mathcal{N}, k \in \mathcal{K}, C_{n,k}(t)\in \mathcal{M}$. The channel process is defined as ${\{C(t)\}}_{t=1}^{\infty}$ with the state space $\cal S$.
Note that $\cal S$ is a finite set with $|\mathcal{S}|= (M+1)^{NK}$. We label each element of $\cal S$ by a positive integer index $s \in \{1,2, ... ,(M+1)^{NK}\}$. Suppose that the channel state matrix associated to the channel state $s$ is denoted by $C_{s}$. The channel state of the system is assumed to have stationary distribution, with stationary probabilities 
$\pi_s=\Pr(C(t)=C_s)$.

Assume that the arrival process to each queue $n \in \cal N$ at time slot $t$ (i.e., the number of packet arrivals during time slot $t$) is represented by $A_n(t)$. Thus, the arrival vector at time slot $t$ is denoted by $A(t)=(A_1(t),A_2(t),...,A_N(t))$. Assume that $E[A_n^2(t)]< A_{max}^2<\infty$ \textit{for all} $t$. Each queue has an infinite buffer space. New arrivals are added to each queue at the end of each time slot. Let $X(t)=(X_1(t), ... , X_N(t))$ be the queue length process vector at the end of time slot $t$ after adding new arrivals to the queues.

A server scheduling policy at each time slot should decide on how to allocate servers (from set $\cal K$) to the queues (in set \(\cal N\)). This must be accomplished based on the available information about the channel state of the system at time slot $t$ (which is $C(t)$) and also the queue length process at the beginning of time slot $t$ (which is $X(t-1)$). In other words, at each time slot $t$, the scheduler has to determine an allocation matrix $I(t) \in \mathcal{I}$ where
\[\mathcal{I}=\left\{I_{N\times K}=\left(I_{n,k}\right),I_{n,k}\in \{0,1\} \mid ~ \sum_{n=1}^N I_{n,k}=1 ~\forall k\in \mathcal{K} \right\} .\]

We can easily observe that the queue length process evolves with time according to the following rule.
\begin{equation}
X^{\mathrm{T}}(t)=\left(X^{\mathrm{T}}(t-1)-\left( C(t)\circledast I(t) \right)\underline{1}_K^{\mathrm{T}} \right)^+ +A^{\mathrm{T}}(t)
\end{equation}
where $(\cdot)^+$ is defined as follows: For an arbitrary vector $v$ of size $|v|$, $(v)^+$ is a vector of the same size whose $i$'th element $(v)^+_i=v_i$  if $v_i\geq 0$ and zero otherwise.
\vspace{-2mm}
\section{Background} \label{back}

\subsection{Strong Stability}     
We begin with introducing the definition of strong stability in queueing systems \cite{Now, phd,power}. Other definitions can be found in \cite{Asmussen, Tassiulas92, Tassiulas93}.
Consider a discrete time single queue system with an arrival process \(A(t)\) and service process $\mu (t)$. Assume that the arrivals are added to the system at the end of each time slot. Hence, the queue length process $X(t)$ evolves with time according to the following rule.
\begin{equation}
X(t)=(X(t-1)-\mu(t))^+ +A(t)
\end{equation}
Strong stability is given by the following definition \cite{Now}.
\begin{defin} 
A queue satisfying the conditions above is called \emph{strongly stable} if 
\begin{equation}
\limsup_{t \to \infty} {1\over{t}} \displaystyle\sum _ {\tau =0}^{t-1}E[X(\tau)] < \infty
\end{equation}   
\end{defin} 

Naturally for a queueing system we have the following definition \cite{Now}.
\begin{defin}  
A queueing system is called to be strongly stable if all the queues in the system are strongly stable. 
\end{defin} 

In this paper, we use the strong stability definition and from now we use ``stability" and ``strong stability" interchangeably.
%
\vspace{-6mm}
\subsection{Some Fundamental Concepts of Polytopes}

In this part, we will have a brief review on the notion of convex polytopes and some fundamental properties of them. These concepts are needed when describing the stability region of MQMS system in section \ref{sec-asli}.
                                                                                                                                                                                                                                                                                                                                                                                                                                                                               
%
%

\begin{defin} 
\label{polyt}
A convex polytope is defined as the convex hull of a finite set of points \cite{lee,alex}.
\end{defin}
%

According to the \textit{Weyl's Theorem} \cite{lee}, a polytope in $\mathbb{R}^N$ always can be expressed by a set $\left\{x \in \mathbb{R}^N  \mid \alpha_{\ell} x^{\mathrm{T}}\leq \beta_{\ell}~,~\ell=1,2,...,L  \right\}$ for some positive integer $L$ and $\alpha_{\ell} \in \mathbb{R}^N$ and $\beta_{\ell} \in \mathbb{R}$.


\textit{Dimension Theorem} \cite{lee}: For a polytope $\mathcal{P} \subset \mathbb{R}^N$, dimension of $\cal P$ is equal to $N$ minus the maximum number of linearly independent equations satisfied by all the points in $\cal P$. 
\begin{defin}
An equality $ax^{\mathrm{T}}\leq b$ is called \textit{valid} for polytope $\mathcal P$ if for every point $x_0 \in \mathcal P$, $ax_0^{\mathrm{T}}\leq b$.
\end{defin}
\begin{defin}
A \textit{face} of polytope $\mathcal P$ is defined as $\mathcal{F}=\{x \in \mathcal{P} \mid ax^{\mathrm{T}}= b\}$ where 
inequality $ax^{\mathrm{T}}\leq b$ is a valid inequality for $\mathcal P$.
\end{defin}

We call the hyperplane $ax_0^{\mathrm{T}}= b$ associated to the valid inequality $ax_0^{\mathrm{T}}\leq b$ a \textit{face defining hyperplane} for $\mathcal{P}$ if its associated face is not empty. In other words, $ax_0^{\mathrm{T}}= b$ is a face defining hyperplane for $\mathcal{P}$ if it intersects with $\mathcal{P}$ on at least one point.
Note that $\mathcal{P}$ has finitely many faces. However, the face defining hyperplanes of a polytope can be infinite.
\begin{defin} 
\label{facet}
A \textit{facet} of polytope $\mathcal P$ is a maximal face distinct from $\cal P$ \cite{alex}. In other words, all faces of $\mathcal{P}$ with dimension $\dim (\mathcal{P})-1$ are called facets of $\mathcal{P}$.
\end{defin}

For polytope $\mathcal{P}=\left\{x \in \mathbb{R}^N  \mid \alpha_{\ell} x^{\mathrm{T}}\leq \beta_{\ell}~,~\ell=1,2,...,L  \right\}$ an inequality is \textit{redundant} if polytope $\cal P$ is unchanged by removing the inequality.

\textit{Redundancy Theorem in Polytopes}: Face defining hyperplanes (inequalities) describing faces of dimension less than $\dim (\mathcal{P})-1$ are redundant \cite{lee}.

Redundancy theorem states that to describe a polytope completely, only facet defining hyperplanes (inequalities) are sufficient. 
\vspace{-2mm}
\section{Necessary and Sufficient Conditions for the Stability of MQMS System} \label{sec-asli}
\subsection{Necessary Conditions for the Stability of MQMS System}
We introduce the departure matrix $H_{N\times K}(t)=(H_{n,k}(t)), n\in {\cal N}, k\in{\cal K}$ in which $H_{n,k}(t)$ represents the total number of packets served by server $k$ from queue $n$ at time slot $t$. Obviously $H_{n,k}(t)\leq C_{n,k}(t)$. 
Thus, the departure process from queue $n$ at time slot $t$ would be $\sum_{k=1}^K H_{n,k}(t)$.
The following equality illustrates the evolution of queue length process with time.
\begin{eqnarray} 
\label{evol} X_n(t)=X_n(t-1)-\displaystyle\sum _{k=1}^K H_{n,k}(t) +A_n(t)~~~\forall n\in \cal N
\end{eqnarray}

Let $\cal G$ be the class of deterministic policies in which at each state of the system, each policy $g\in \cal G$ allocates the servers according to a predetermined allocation matrix. Therefore, each deterministic policy $g$ is specified by $|\cal S|$ allocation matrices $I_s^{(g)}, ~s\in \cal S$. 
 Note that set $\cal I$ is a finite set and since the channel state space is also finite, set $\cal G$ is finite.

A deterministic policy $g$ will result into an average transmission rate for each queue $n$. Let $R_n^{(g)}$ denote the time averaged transmission rate provided to queue $n$ and $R^{(g)}:=(R_1^{(g)},R_2^{(g)},...,R_N^{(g)})$.
It follows that,\vspace{-.5mm}
\begin{eqnarray}
\label{rg}
R^{(g)}={\left(\displaystyle\sum_{s\in {\cal S}}\pi_s \left(C_s\circledast I_s^{(g)} \right)\underline{1}_K^{\mathrm{T}}\right)}^{\mathrm{T}}
\end{eqnarray}   

Each rate vector $R^{(g)}$ determines a single point in $\mathbb{R}_{+}^N$. Now, consider the convex hull of all the points $R^{(g)}$, $g \in \cal G$ in $\mathbb{R}_{+}^N$, i.e., $\mathcal{R}=conv.hull(R^{(g)})$.  

According to the above discussion and definition of convex polytope, we can show that the set of achievable transmission rate vectors $\cal R$ is specified by a polytope. Henceforth, we denote the achievable transmission rate polytope by $\cal P$.

Note that convex hull representation of the stability region may not be suitable to use in network optimization problems like (\ref{fairness}). Since it does not provide mathematical statements in a form to be used as the constraints in (\ref{fairness}). In the following, we will characterize the stability region by a set of linear inequalities.
To introduce the necessary conditions for the stability of MQMS system we need to prove a series of Lemmas (Lemmas \ref{l1} to \ref{l5}). In this paper, the proofs were skipped due to space limitation.

\begin{lem}
\label{l1}If the MQMS system is strongly stable under a server allocation policy \(\Delta\), then for any vector $\alpha=(\alpha_1,\alpha_2,...,\alpha_N)\in \mathbb{R}_{+}^N$ we have \vspace{-1mm}
\begin{eqnarray}
\label{l1_1}\lim_{t\rightarrow \infty}\frac{1}{t} \displaystyle\sum_{\tau=1}^{t} \alpha E[A(\tau)]^{\mathrm{T}}=\lim_{t\rightarrow \infty} \frac{1}{t}\displaystyle\sum_{\tau=1}^{t} \alpha E[H(\tau)]\underline{1}_K^{\mathrm{T}}.
\end{eqnarray}
\end{lem}

%

 

\begin{lem}
\label{l3}
If the MQMS system is strongly stable under a server allocation policy \(\Delta\), then  $\overline{E[A(t)]} \in \cal P$.
\end{lem}
\begin{lem}
\label{l4}
If the MQMS system is strongly stable under a server allocation policy \(\Delta\), then \textit{for all} $\alpha \in \mathbb{R}_+^N$
\begin{eqnarray} 
\label{l4_1} 
\alpha \overline{E[A(t)]^{\mathrm{T}}} \leq \displaystyle\sum_{s\in \mathcal{S}}\pi_s 
  \max_{I \in \mathcal I}  \left(\alpha (C_s \circledast I )\underline{1}_K^{\mathrm{T}} \right)
\end{eqnarray}
\end{lem}
\vspace{-1mm}


As we can see, the number of inequalities defined in (\ref{l4_1}) is infinite. Note that each inequality in (\ref{l4_1}) determines a valid inequality for polytope $\cal P$. It is not hard to show that the hyperplanes associated to the valid inequalities of (\ref{l4_1}) are all face defining hyperplanes of polytope $\cal P$. To see this, let ${\cal I}^{\alpha}$ denote a set of allocation matrices of $\{I^{\alpha}_s, s\in \cal S \}$ that maximizes the right hand side of (\ref{l4_1}), i.e.,
\begin{eqnarray}
\label{argmax}I_s^{\alpha} = \arg\max_{I \in \cal I} \alpha (C_s \circledast I )\underline{1}_K^{\mathrm{T}}.
\end{eqnarray}
Note that ${\cal I}^{\alpha}$ is not unique and there may be more than one set of allocation matrices of ${\cal I}^{\alpha}$ whose elements maximize  $\alpha (C_s \circledast I )\underline{1}_K^{\mathrm{T}}$.
According to (\ref{rg}) and definition of polytope $\cal P$, ${\left(\sum_{s\in \mathcal{S}}\pi_s 
\left( C_s \circledast I_s^{\alpha} \right) \underline{1}_K^{\mathrm{T}}\right)}^{\mathrm{T}} \in \cal P$. On the other hand,  
$
\sum_{s\in \mathcal{S}}\pi_s 
 \alpha (C_s \circledast I^{\alpha}_s)\underline{1}_K^{\mathrm{T}} =\alpha \sum_{s\in \mathcal{S}}\pi_s 
  (C_s \circledast I^{\alpha}_s )\underline{1}_K^{\mathrm{T}}
$. 
Thus, ${\left(\sum_{s\in \mathcal{S}}\pi_s 
\left( C_s \circledast I_s^{\alpha} \right) \underline{1}_K^{\mathrm{T}}\right)}^{\mathrm{T}}$ is located on the hyperplane associated to (\ref{l4_1}).
In other words, the hyperplane associated to (\ref{l4_1}) touches polytope $\cal P$.
So, the set of inequalities of (\ref{l4_1}) are all face defining hyperplanes of polytope $\cal P$.
According to redundancy theorem in polytopes \cite{lee}, not all the face defining hyperplanes of a polytope are required to determine a polytope completely. The inequalities corresponding to the facets of a polytope are just sufficient to characterize a polytope. In the following, we will characterize the facets of polytope $\cal P$.

Let $V$ denote the set of vectors $\alpha \in \mathbb{R}^N_+$ with the following property, i.e.,
\begin{eqnarray} \label{Vdef}
\lefteqn{V:=\left\{ \alpha \in \mathbb R^N_+ \mid \right. } \notag\\ 
&& ~~\forall~ ( \mathcal{U} \subset{\cal N}, \mathcal{U}\neq\varnothing, \alpha_{\mathcal{U}}\neq \underline{0}_{|\mathcal{U}|}, \alpha_{\mathcal{U}^c}\neq \underline{0}_{|\mathcal{U}^c|}) \nonumber\\&& 
 ~\left. ~ \exists ~( i \in  \mathcal{U} , j \in  \mathcal{U}^c ,~ m,n \in \mathcal M ,~ \alpha_{i}, \alpha_{j}, m,n\neq 0 ) \right. \nonumber \\
 && ~~~~~~~~~~~~~~~~~~~~~~~~~~~~~~~~~~~~~~~~~~ \left. : \alpha_{i}m=\alpha_{j}n  \right\}~~~~~~
\end{eqnarray}

In other words, a vector $\alpha$ belongs to set $V$ if for any partitioning of elements of vector $\alpha$ into two non-empty disjoint (sub)vectors in which all the elements of each vector are not identically equal to zero, there exists at least one non-zero element in each vector, the ratio of which is equal to ratio of two non-zero elements of set $\cal M$. To clarify the definition of set $V$ consider the following example.

\el
 Let ${\cal M}=\{0,1,2\}$ and $N=4$. Consider vector $\alpha_1=(1,2, 5 ,10)$. We can partition $\alpha_1$ to $\alpha_{1_{\{1,2\}}}=(1,2)$ and $\alpha_{1_{\{3,4\}}}=(5,10)$. Note that there exist no elements in $\alpha_{1_{\{1,2\}}}$ and $\alpha_{1_{\{3,4\}}}$ whose ratio is 1 or 2 and therefore, $\alpha_1 \notin V$. Another example is the vector $\alpha_2=(0,2.5,0,0)$ in which for any partitioning of the vector into two non-empty disjoint parts, one part will always have all the elements identically equal to zero. Therefore, $\alpha_2 \in V$.

In the following we will show that any vector in ${\mathbb R^N_+}-V$ cannot form a facet defining hyperplane for polytope $\cal P$ and therefore the inequalities derived by the vectors of set ${\mathbb R^N_+}-V$ are redundant to characterize polytope $\cal P$.
\begin{lem} \label{l6}
 The hyperplane associated to the valid inequality of (\ref{l4_1}) is not a facet defining hyperplane of $\cal P$ if $\alpha \in {\mathbb R^N_+}-V$.
\end{lem}

Set $V$ is an infinite set. However, we can show that set $V$ is finite up to multiplication of vectors by positive scalars. To show this, we define set ${\cal W}=\left\{z \in \mathbb{Z_+} \mid  z=\prod_{j=1}^{N-1}m_j,~m_j \in \cal M \right\}$. Then, we consider the set $\mathcal{W}^N=\{(\alpha_1,\alpha_2,...,\alpha_N)  \mid \alpha_n \in \mathcal{W}\} $. We can prove the following Lemma.
\begin{lem}
\label{l5}
There exists $\widehat{V} \subseteq {\cal W}^N$ such that any vector $\alpha \in V$ can be written as $\alpha=q \beta$ for some $\beta \in \widehat{V}$ and $q>0$.

\end{lem}


We can show that $|\mathcal{W}^N|= \left( {M+N-2 \choose{N-1}}+1 \right)^N $. Since $\widehat{V} \subseteq {\cal W}^N $, therefore 
$|\widehat{V}| \leq |\mathcal{W}^N|-1$ (excluding the vector $\underline{0}_N$).

We now introduce the necessary conditions for the stability of MQMS system with stationary channel distributions by the following theorem.
\begin{theor} 
\label{t1}
 If there exists a server allocation policy \(\triangle\) under which the system is stable, then 
\begin{eqnarray}
\label {t1_1} \alpha \overline{E[A(t)]^{\mathrm{T}}} \leq \displaystyle\sum_{s\in \mathcal{S}}\pi_s 
  \max_{I \in \mathcal I}  \left(\alpha (C_s \circledast I )\underline{1}_K^{\mathrm{T}} \right),~~ \alpha \in \widehat{V}.
\end{eqnarray}
\end{theor}
\vspace{-2mm}The proof follows from Lemmas \ref{l1} to \ref{l5}. 

Note that according to Theorem \ref{t1}, in order to characterize polytope $\cal P$, we have to specify all the elements of set $\widehat{V}$ which can be obtained after considering all possible vectors $\alpha \in  \mathcal{W}^N- \underline{0}_N$ and removing redundancies following (\ref{Vdef}). This can be done using numerical computations. We may also use $\alpha \in  \mathcal{W}^N- \underline{0}_N$ instead of $\alpha \in \widehat{V}$ in (\ref{t1_1}).
%
Although by using $ \mathcal{W}^N- \underline{0}_N $ we may obtain some redundant inequalities, since $|\mathcal{W}^N- \underline{0}_N|< \infty$ we still have a finite number of inequalities to describe polytope $\mathcal{P}$. 
Table \ref{table} depicts $\widehat{V}$ and $|\mathcal{W}^N|-1$ for some simple cases of $N=2,3$ and $M=1,2,3,4$. \vspace{-2mm}
\begin{table}[h]
\renewcommand{\arraystretch}{1.3}
\caption{$|\widehat{V}|$ and $|\mathcal{W}^N|-1$ for $N=2,3$ and $M=1,2,3,4$}
\label{table}
\centering 
\begin{tabular}{cc|c|c|c|c|l}
\cline{3-6} 
& & $M=1$ & $M=2$ & $M=3$ & $M=4$ \\ \hline
\cline{1-6} 
\multicolumn{1}{|c||}{\multirow{2}{*}{\hspace{-1mm}$|\widehat{V}|,|\mathcal{W}^N|-1 $\hspace{-1mm}}} &
\multicolumn{1}{|c|} {$N=2$} & $3,3$ & $5,8$ & $9,15$ &$12,24$     \\ \cline{2-6}
\multicolumn{1}{|c||}{}                        &
\multicolumn{1}{|c|}{$N=3$} & \hspace{-.5mm}$7,7$\hspace{-.5mm} & $25,63$ & $109,342$ & \hspace{-.5mm}$253,1330$\hspace{-.5mm}  \\
%
\hline
\end{tabular}
\end{table}

\begin{cor} \label{nec-stat}
 For an MQMS system with stationary arrival processes, we have $E[A(t)]=\lambda=(\lambda_1,\lambda_2,...,\lambda_N)$ and the necessary conditions for the stability of the system would be
\begin{eqnarray}
\label {t1_2}   \alpha {\lambda}^{\mathrm{T}} \leq \displaystyle\sum_{s\in \mathcal{S}}\pi_s 
  \max_{I \in \mathcal I}  \left(\alpha (C_s \circledast I )\underline{1}_K^{\mathrm{T}} \right),~~ \alpha \in \widehat{V},
\end{eqnarray}
\end{cor}


\begin{cor}
 For an MQMS system with ON-OFF channels, we have ${\cal W}=\{0,1\}$ and therefore the necessary conditions for the stability of the system become
\begin{eqnarray}
\label {t1_3} \alpha \overline{E[A(t)]^{\mathrm{T}}} \leq \displaystyle\sum_{s\in \mathcal{S}}\pi_s 
  \max_{I \in \mathcal I}  \left(\alpha (C_s \circledast I )\underline{1}_K^{\mathrm{T}} \right),~ \alpha_n \in \{0,1\}  
\end{eqnarray}
In a special case considered in \cite{khodam} where the channels are modelled by independent Bernoulli random variables with $ E[C_{n,k}(t)]=p_{n,k}$, the necessary conditions for the stability of the system are given by \vspace{-3mm}
 \begin{eqnarray}
\label {t1_4} 
\overline{\displaystyle\sum_{n\in Q}E[A_n(\tau)]}\leq K-\displaystyle\sum_{k=1}^{K}\displaystyle\prod_{n\in Q}(1-p_{n,k})~~ \forall Q\subseteq \mathcal{N}.
\end{eqnarray}
 In this case, the total number of inequalities needed to describe the polytope $\cal P$ is equal to $2^N-1$.
\end{cor}

\vspace{-2mm}
\subsection{Sufficient Conditions for the Stability of MQMS System}
 Consider a server allocation policy which determines the allocation matrix at each time slot $t$ by solving the following maximization problem.
\begin{eqnarray}
\label{mw}
I(t) = \arg\max_{I \in \cal I} X(t-1) (C(t) \circledast I )\underline{1}_K^{\mathrm{T}}
\end{eqnarray}
This policy is called Maximum Weight (MW) and was first introduced in \cite{javidi2,Tassiulas92}. According to the constraints on allocation matrix $I$, we can easily conclude that MW policy allocates the servers by the following rule:
At each time slot $t$, MW policy allocates each server $k$ to queue $n$ who has the maximum $X_n(t-1)C_{n,k}(t)$. 
In the special case where the channels are ON-OFF, the policy is equivalent to AS/LCQ (Any Server/Longest Connected Queue) introduced in \cite{khodam}. In the following, we derive sufficient conditions for the stability of our model and prove that MW stabilizes the system as long as condition (\ref{sc1}) is satisfied. An upper bound is also derived for the time averaged expected number of packets in the system.
\begin{theor}\label{sc} The MQMS system is stable under MW if \emph{for all} \(t\) 
\begin{equation}
\label{sc1}\alpha E[A(t)]^{\mathrm{T}} < \displaystyle\sum_{s\in \mathcal{S}}\pi_s 
  \max_{I \in \mathcal I}  \left(\alpha (C_s \circledast I )\underline{1}_K^{\mathrm{T}} \right),~~ \alpha \in \widehat{V}
\end{equation}
Furthermore, the following bound for the average expected ``aggregate" occupancy holds.
\begin{eqnarray}
\label{sc2} \limsup_{t \to \infty} {1\over{t}} \displaystyle\sum_ {\tau =0}^{t-1} \displaystyle\sum_{n=1}^N E[X_n(\tau)]\leq \frac{N A_{max}^2+(MK)^2}{2\delta} 
\end{eqnarray}
where $\delta$ is the maximum positive number such that \textit{for all t} we have $E[A(t)]+\delta \underline{1}_N \in  \mathcal{P}$.
%
\end{theor}

\cor \label{suf-stat}
 For an MQMS system with stationary arrival processes, we have $E[A(t)]=\lambda$ and the sufficient conditions for the stability of the system under MW policy become $\alpha {\lambda}^{\mathrm{T}} < \sum_{s\in \mathcal{S}}\pi_s  \max_{I \in \mathcal I}  \left(\alpha (C_s \circledast I )\underline{1}_K^{\mathrm{T}} \right),~~ \alpha \in \widehat{V}$.

According to Corollaries \ref{nec-stat} and \ref{suf-stat} and the definition of network capacity region, we can conclude that for an MQMS system with stationary channel distribution and stationary arrival processes, the stability region is characterized by (\ref{t1_2}). 

%

Consider an MQMS system with ON-OFF channels. For such a system, the MW policy is equivalent to AS/LCQ policy introduced in \cite{khodam}. In AS/LCQ the policy takes an arbitrary ordering of servers (to be allocated to the queues) and then for each server, AS/LCQ allocates it to its longest connected queue (LCQ). 
Note that in such a system, ${\cal W}=\{0,1\}$ and therefore we will have the following corollary.

\begin{cor}
The MQMS system with ON-OFF channels is stable under AS/LCQ if \textit{for all} $t$
\begin{eqnarray}
\label {t2_3} 
 \alpha E[A(t)]^{\mathrm{T}} < \displaystyle\sum_{s\in \mathcal{S}}\pi_s 
  \max_{I \in \mathcal I}  \left(\alpha (C_s \circledast I )\underline{1}_K^{\mathrm{T}} \right),~~ \alpha_n \in \{0,1\} \vspace{-1mm}
\end{eqnarray}
In a special case where the channels are modelled by independent Bernoulli random variables with $ E[C_{n,k}(t)]=p_{n,k}$, AS/LCQ stabilizes the system if \vspace{-2mm}
\begin{eqnarray}
\label {t2_4} 
\displaystyle\sum_{n\in Q}E[A_n(\tau)] < K-\displaystyle\sum_{k=1}^{K}\displaystyle\prod_{n\in Q}(1-p_{n,k})~~ \forall Q\subseteq \mathcal{N}
\end{eqnarray}
\end{cor}

\vspace{-4mm}
\subsection{Stability Region for Fluid Model MQMS Systems with Stationary Continuous Channel Distributions}

We will consider a time slotted fluid model MQMS system with stationary channel distributions. 
In this case, the amount of work that arrives into (and departs from) the queues is considered to be a continuous process. We also assume that the channel state of the link from each queue to each server is modelled by a continuous random variable.
The channel state matrix  in the fluid model MQMS system is defined as $C(t)=\left( C_{n,k}(t)\right),n\in \mathcal{N}, k \in \mathcal{K}, C_{n,k}(t)\in \mathbb{R}_+$.
We assume that the channel state follows a stationary distribution $f_C(c)$. $f_C$ is the joint distributions of all $C_{n,k}$ variables.
We can easily check that in the fluid model MQMS system, set $\widehat{V}$ is equivalent to $\mathbb{R}_+^N$ as in the fluid model MQMS system, set $\cal M$ is replaced with $\mathbb{R}_+$. Therefore, stability region for the fluid model system is characterized by the following set of inequalities.
\begin{eqnarray}
\label {cont}
\lefteqn{ \alpha {\lambda}^{\mathrm{T}}  \leq \int_{c_{n,k}=0}^{\infty}\int_{c_{n,k-1}=0}^{\infty}  \cdots \int_{c_{1,1}=0}^{\infty}    
  \max_{I \in \mathcal I}  \left(\alpha (c \circledast I )\underline{1}_K^{\mathrm{T}} \right)}
\notag ~~~~~~~~~\\  
 && \times f_C(c)~dc_{1,1}~\cdots ~dc_{n,k-1}~dc_{n,k} ~~ \alpha \in \mathbb{R}_+^N~~~~~~ 
\end{eqnarray}

Note that in order to characterize the stability region of the fluid model system we need to compute an infinite number of integrals in (\ref{cont}). In fact, the stability region of fluid model system is characterized by infinite number of half spaces which means that the stability region is a \textit{convex surface}. In this case, depending on the channel distribution and dimension of the system, we could characterize the stability region by a limited number of non-linear inequalities instead of infinite number of linear inequalities as we show in the following example.

\el
Consider a fluid model MQMS system with two queues and one server. Suppose that the channel state variables $C_{1,1}$ and $C_{2,1}$ are independent and follow exponential distribution with means $\mu_1$ and $\mu_2$, respectively. Such a modelling is used to model slow Rayleigh fading channels with low SNR where the approximation $\log(1+x)\simeq x$ is valid for small positive $x$. We can show that the stability region for this example is described by 

\begin{figure}[tp]
    \centering
    \includegraphics[width=0.5\textwidth]{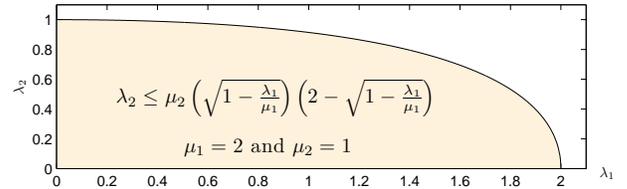}\vspace{-1mm}
	\caption{ Stability region for $\mu_1=2$ and $\mu_2 =1$}
	\label{cont-2q}\vspace{-4mm}
\end{figure}

\vspace{-5mm}
\begin{eqnarray}\vspace{-5mm}
\label{stab-2q}\lambda_2 \leq \mu_2 \left( \sqrt{1-\frac{\lambda_1}{\mu_1}}\right) \left( 2-\sqrt{1-\frac{\lambda_1}{\mu_1}}  \right)
\end{eqnarray}
for $\lambda_1 \geq 0$ and $\lambda_2 \geq 0$. The derivation of (\ref{stab-2q}) is skipped because of space limitation.  
As we can see from equation (\ref{stab-2q}), the stability region in this example is characterized by just one non-linear inequality (\ref{stab-2q}) and two linear inequalities $\lambda_1 \geq 0$ and $\lambda_2 \geq 0$.
The stability region for this example for $\mu_1=2$ and $\mu_2 =1$ is illustrated in Figure \ref{cont-2q}.

\vspace{-5mm}
\section{Conclusions}\label{conc}
In this paper, we characterized the stability region polytope of multi-queue multi-server (MQMS) queueing systems with stationary channel and packet arrival processes by a finite set of linear inequalities given by (\ref{t1_2}). 
 We also argued that in general it may be (computationally) hard to quantify set $\widehat{V}$ from set $\mathcal{W}^N$. In this case, we can use set $\mathcal{W}^N$ itself instead of $\widehat{V}$ in (\ref{t1_2}) as $\widehat{V} \subseteq \mathcal{W}^N$, although we may have some redundant inequalities. 
We also considered the stability region for a fluid model MQMS. In this case, we determine the stability region by an infinite set of linear inequalities given by (\ref{cont}). Finally, by use of an example we showed that depending on the channel distribution and number of queues, we may characterize the stability region by a finite set of non-linear inequalities instead of infinite set of linear inequalities.
\vspace{-1mm}
\bibliographystyle{ieeetran}

\bibliography{Ref}

\end{document}